\documentclass[11pt]{amsart}

\usepackage{amscd}
\usepackage{amsmath}
\usepackage{graphicx}
\usepackage{amsfonts}
\usepackage{amssymb}
\begin{document}
\title[Constructing entanglement witnesses]{Constructing entanglement witnesses for infinite-dimensional systems}

\author{Jinchuan Hou}
\address[Jinchuan Hou]{Department of Mathematics, Taiyuan
University of Technology, Taiyuan 030024, P. R. of China; Department
of Mathematics, Shanxi University, Taiyuan, 030006, P R. China}
\email{jinchuanhou@yahoo.com.cn}

\author{Xiaofei Qi}
\address[Xiaofei Qi]{
Department of Mathematics, Shanxi University, Taiyuan 030006, P.R.
China.} \email{qixf1980@126.com}

\thanks{{\it This work is partially supported by National
Natural Science Foundation of China (No. 10771157), Research Grant
to Returned Scholars of Shanxi (2007-38) and Foundation of Shanxi
University.}}

\begin{abstract}

It is shown that, every entangled state in an infinite-dimensional
composite system has a simple entanglement witness of the form
$\alpha I+T$ with $\alpha$ a nonnegative number and $T$ a finite
rank self-adjoint operator. We also provide two methods of
constructing entanglement witness and apply them to obtain some
entangled states that cannot be detected by the PPT criterion and
the realignment criterion.

\end{abstract}
\maketitle

\section{Introduction}

In quantum mechanics, a quantum system is
 associated with a separable complex Hilbert space $H$, i.e.,
the state space.  A quantum state is described as a density operator
$\rho\in{\mathcal B}(H)$ (namely, $\rho$ is positive and has trace
1); furthermore, $\rho$ is a pure state if $\rho^2=\rho$; $\rho$ is
a mixed state if $\rho^2\not=\rho$. The state space $H$ of a
composite quantum system is a tensor product of several state spaces
$H_i$, that is $H=H_1\otimes H_2\otimes \ldots \otimes H_k$. Let $H$
and $K$ be finite dimensional and let $\rho$ be a state acting on
$H\otimes K$. $\rho$ is said to be separable if $\rho$ can be
written as
$$\rho=\sum_{i=1}^k p_i \rho_i\otimes \sigma _i,$$
where $\rho_i$ and $\sigma_i$ are states on $H$ and $K$
respectively, and $p_i$ are positive numbers with $\sum
_{i=1}^kp_i=1$. Otherwise, $\rho$ is said to be inseparable or
entangled (ref. \cite{BZ, NC}). For the case that at least one of
$H$ and $K$ is of infinite dimension,  by R. F. Werner \cite{W},
{\it a state  $\rho$ acting on $H\otimes K$ is called separable if
it can be approximated in the trace norm by the states of the form
$$\sigma=\sum_{i=1}^n p_i \rho_i\otimes \sigma _i,$$
where $\rho_i$ and $\sigma_i$ are states on $H$ and $K$
respectively, and $p_i$ are positive numbers with
$\sum_{i=1}^np_i=1$}. \if ; or equivalently, $\rho=\sum_{i=1}^\infty
p_i \rho_i\otimes \sigma _i,$ with $p_i\geq 0$ and
$\sum_{i=1}^\infty p_i=1$.\fi Otherwise, $\rho$ is called an
entangled state.

Entanglement is of special significance in quantum information
processing and is responsible for many quantum tasks such as
teleportation, dense coding, key distribution, error correction etc.
(see \cite{NC,BB,BP,DE,DE1,S}). It is very important but also
difficult to determine whether or not a state in a composite system
is separable. For the case of $2\times 2$ or $2\times 3$ systems,
that is, for the case $\dim H=\dim K=2$ or $\dim H=2, \ \dim K=3$, a
state is separable if and only if it is PPT (Positive Partial
Transpose) \cite{Hor, Pe}. But PPT is only a necessary condition for
a state  to be separable acting on Hilbert space of higher
dimensions. There are PPT states that are entangled. It is known
that PPT entangled states belong to the class of bound entangled
states \cite{HHH}. In \cite{CW}, the realignment criterion for
separability in finite dimensional systems was found. It is
independent of the PPT criterion and  can detect some bound
entangled states that cannot be recognized by the PPT criterion.

A most general approach to study the entanglement of quantum states
in finite dimensional physical systems is based on the notion of
entanglement witnesses (see \cite{Hor}). A Hermitian operator $W$
acting on $H\otimes K$ is said to be an entanglement witness
(briefly, EW), if  $W$ is not positive and ${\rm Tr}(W\sigma)\geq 0$
holds for all separable states $\sigma$. Thus, if $W$ is an EW, then
there exists an entangled state $\rho$ such that ${\rm Tr}(W\rho) <
0$ (that is, the entanglement of $\rho$ can be detected by $W$). It
was shown that, a state is entangled if and only if it is detected
by some entanglement witnesses \cite{Hor}. This entanglement witness
criterion is also valid for infinite dimensional systems. Clearly,
constructing entanglement witnesses is a hard task. There was a
considerable effort in constructing and analyzing the structure of
entanglement witnesses \cite{B,TG,CK,JB} (see also \cite{HHH1} for a
review). However, few literature concerns the structure properties
of entanglement witnesses for infinite dimensional composite systems
since the witnesses for infinite dimensional case have  much more
complicated structure. A general discussion can be found in
\cite{SV1}, there it was shown that every entanglement witness $W$
in a bipartite system can be written as of the form $W=cI-C$ with
$I$ the identity operator, $C$ a positive operator and $c$ a
positive number (this is obvious by taking $c=\|W\|$ and
$C=\|W\|I-W$).

In the present paper we continue discussing the structure properties
of entanglement witnesses for infinite dimensional composite
systems. The main purpose is to answer the question, for any
entangled state, whether or not we can find some witnesses of simple
structure that detect the given state. We show that a state $\rho$
on an infinite-dimensional composite system is entangled if and only
if there exists an entanglement witness of the form $W=\alpha I+T$
with $\alpha$ a nonnegative number and $T$ a
 finite rank self-adjoint operator (i.e., the  range of $T$ is finite-dimensional) such that ${\rm Tr}(W\rho)<0$.
 This is remarkable because  such kind of witness $W$ is a Fredholm
 operator of index 0 (i.e., $W$ has a closed range and ind$(W)=\dim\ker( W)-\dim\ker
 (W^*)=0$) and the spectrum of $W$
 consists of finite many eigenvalues.
 We also
develop  two methods of constructing such entanglement witnesses,
which are applied to find entanglement witnesses of such form that
can detect some entangled states, however these states cannot be
detected by the PPT criterion and the realignment criterion.

Throughout, $H$ and $K$ are separable complex Hilbert spaces that
may be of infinite dimension  if no specific assumption is made, and
$\langle \cdot|\cdot\rangle$ stands for the inner product in both of
them. ${\mathcal B}(H,K)$ (${\mathcal B}(H)$ when $K=H$) is the
Banach space of all (bounded linear) operators from $H$ into $K$.
$A\in{\mathcal B}(H)$ is self-adjoint if $A=A^\dagger$ ($A^\dagger$
stands for the adjoint operator of $A$); and $A$ is positive,
denoted by $A\geq 0$, if $A$ is self-adjoint with the spectrum
falling in the interval $[0,\infty)$ (or equivalently, $\langle \psi
| A|\psi\rangle\geq 0$ for all $|\psi\rangle\in H$). For
$A\in{\mathcal B}(H)$, denote $|A|=(A^\dagger A)^{\frac{1}{2}}$.
Then the trace class ${\mathcal T}(H)=\{T : \|T\|_1={\rm
Tr}(|T|)<\infty\}$ and the Hilbert-Schmidt class ${\mathcal
{HS}}(H)=\{T : \|T\|_2=({\rm Tr}(T^\dagger
T))^{\frac{1}{2}}<\infty\}$. The trace class and the Hilbert-Schmidt
class are ideals in ${\mathcal B}(H)$ with ${\mathcal
T}(H)\subseteq{\mathcal {HS}}(H)$. Furthermore, ${\mathcal T}(H)$ is
a Banach space with the trace norm $\|\cdot\|_1$ and ${\mathcal
{HS}}(H)$ is a Hilbert space with inner product $\langle
A,B\rangle={\rm Tr}(B^\dagger A)$. Let ${\mathcal B}_0(H)$ be the
subspace of all compact operators in ${\mathcal B}(H)$. It is a well
known fact that the dual space of ${\mathcal T}(H)$ (resp. of
${\mathcal B}_0(H)$) is ${\mathcal T}(H)^*={\mathcal B}(H)$ (resp.
is ${\mathcal B}_0(H)^*={\mathcal T}(H)$) and every bounded linear
functional is of the form $T\mapsto{\rm Tr}(AT)$, where
$A\in{\mathcal B}(H)$ (resp. $A\in{\mathcal T}(H)$).

\section{Witnesses of  simple form }

Let  $\rho$ be a state on $H\otimes K$. In \cite{Hor}, it was shown
that $\rho$ is entangled if and only if there exists a self-adjoint
operator $W\in{\mathcal B}(H\otimes K)$ such that $W$ is an EW and
${\rm Tr}(W\rho)<0$. However, for infinite-dimensional systems, the
structure of the entanglement witnesses are very complicated and
difficult to deal with. Then it is interesting to ask: for any
entangled state $\rho$, can we always find a witness $W$ detecting
$\rho$ of some simple structure that is easily handled?

In this section we will improve the  result in \cite{Hor} and give
an answer of the question above. Our main result shows that any
entangled state can be detected by an entanglement witness of the
form ``nonnegative constant times identity + a self-adjoint operator
of finite rank ". It is obvious that such kind of witnesses are much
simple and easily handled.  To do this, we first give a lemma.

{\bf Lemma 2.1.} {\it Let $H, K$ be complex Hilbert spaces with at
least one infinite-dimensional, and $\rho$ be a state on $H\otimes
K$. Then $\rho$ is entangled if and only if there exists a real
number $\alpha\in{\mathbb R}$ and a trace-class self-adjoint
operator $T$ such that the operator $W=\alpha I+T$ satisfies
$${\rm Tr}(W\rho)<0 $$and
$${\rm Tr}(W\sigma )\geq 0\ \ \mbox{\rm for \ all \ separable  \ states}\ \ \sigma \ {\rm
on}\  H\otimes K.$$}

{\bf Proof.} Obviously, we only need to prove the ``only if" part by
the result in \cite{Hor}. Denote by ${\mathcal B}_0(H\otimes K)$ the
Banach space of all compact operators on $H\otimes K$. Let
${\mathcal S}_0=\{\sigma=\sum_{i=1}^n p_i\sigma_i\otimes \eta_i:
\sum_{i=1}^np_i=1, \sigma_i\in{\mathcal B}(H), \eta_i\in{\mathcal
B}(K)$ are rank-one projections$\}$. Let ${\mathcal C}=$ the
operator norm closure of ${\mathcal S}_0$ and let ${\mathcal S}_{\rm
sep}$  be the set of all separable states acting on $H\otimes K$,
i.e., the trace-norm closure of ${\mathcal S}_0$. It is obvious that
${\mathcal S}_{\rm sep}\subset{\mathcal C}$. Furthermore, $\mathcal
C$ is a closed convex subset in ${\mathcal B}_0(H\otimes K)$. We
claim that if $\rho$ is an entangled state on $H\otimes K$, then
$\rho\not\in{\mathcal C}$. If, on the contrary, $\rho\in{\mathcal
C}$, then $\lim _{n\rightarrow\infty}\|\rho-\sigma_n\|=0$ for some
$\{\sigma_n\}\subset {\mathcal S}_0$, here we denote by $\|A\|$  the
operator norm for an operator $A$, as usual. Take any orthonormal
bases $\{|i\rangle\}$ and $\{|j\rangle\}$ of $H $ and $K$,
respectively. For any positive integers $k\leq \dim H$ and
$l\leq\dim K$, denote $P_{kl}=P_k\otimes Q_l$, where $P_k$ and $Q_l$
are finite rank projections onto the subspaces spanned by
$\{|i\rangle\}_{i=0}^k$ and $\{|j\rangle\}_{j=0}^l$, respectively.
Since $\rho$ is entangled, by \cite[Theorem 2]{SV2}, there exists
$(k,l)$ such that $\rho_{kl}={\rm Tr}(P_{kl}\rho
P_{kl})^{-1}P_{kl}\rho P_{kl} $ is entangled. It follows from $\lim
_{n\rightarrow\infty}\|\rho-\sigma_n\|=0$ that $\lim
_{n\rightarrow\infty}\|P_{kl}\sigma_nP_{kl}-P_{kl}\rho P_{kl}\|=0$.
Since $P_{kl}\sigma_nP_{kl}$ and $P_{kl}\rho P_{kl}$ can be regarded
as operators acting on the finite-dimensional space $P_kH\otimes
Q_lK$, we must have $\lim_{n\rightarrow\infty}{\rm
Tr}(P_{kl}\sigma_nP_{kl})={\rm Tr}(P_{kl}\rho P_{kl})>0$. Thus for
sufficient large $k$ and $l$, ${\rm
Tr}(P_{kl}\sigma_nP_{kl})\not=0$. Let $\sigma _{nkl}={\rm
Tr}(P_{kl}\sigma_nP_{kl})^{-1}P_{kl}\sigma_nP_{kl}$. Then it is
clear that $\sigma _{nkl}$ is separable and $\lim
_{n\rightarrow\infty}\|\rho_{kl}-\sigma_{nkl}\|=0$. Again, regarding
$\rho_{kl}$ and $\sigma_{nkl}$ as states acting on the
finite-dimensional space $P_kH\otimes Q_lK$, we conclude that $\lim
_{n\rightarrow\infty}\|\rho_{kl}-\sigma_{nkl}\|_1=0$. Therefore,
$\rho_{kl}$ is separable, a contradiction. So, a state $\rho$ is
entangled implies that $\rho\not\in{\mathcal C}$, as desired.

Now, we can apply the Hahn-Banach theorem to $\rho$ and $\mathcal
C$, and get a linear functional $f\in{\mathcal B}_0(H\otimes K)^*$
and a real number $\alpha$ such that Re$f(\rho)<\alpha \leq {\rm
Re}f(C)$ for all $C\in{\mathcal C}$, here Re denotes the real part.
Note that ${\mathcal B}_0(H\otimes K)^*={\mathcal T}(H\otimes K)$
and thus there exists a trace-class operator $B\in{\mathcal
T}(H\otimes K)$ such that $f(X)={\rm Tr}(XB)$ holds for every
$X\in{\mathcal B}_0(H\otimes K)$. Let $T=\frac{1}{2}(B+B^\dagger )$.
Then $T$ is a self-adjoint trace-class operator satisfying ${\rm
Tr}(T\rho )<\alpha\leq{\rm Tr} (TC)$ for all $C\in{\mathcal C}$.
Particularly, ${\rm Tr}(T\rho )<\alpha\leq{\rm Tr} (T\sigma )$ for
all $\sigma\in{\mathcal S}_{\rm sep}$. Now, let $W=T-\alpha I$, then
${\rm Tr}(W\rho )<0\leq{\rm Tr} (W\sigma )$ for all
$\sigma\in{\mathcal S}_{\rm sep}$. \hfill$\square$

Based on Lemma 2.1, we can get the main result in this section.

{\bf Theorem 2.2.} {\it Let $H, K$ be complex Hilbert spaces with at
least one infinite-dimensional, and $\rho$ be a state on $H\otimes
K$. Then $\rho$ is entangled if and only if there exists some real
number  $\alpha\geq 0$ and some finite rank self-adjoint operator
$R$ on $H\otimes K$ such that the operator $W=\alpha I+R$
 satisfies
$${\rm Tr}(W\rho)<0 $$and
$${\rm Tr}(W\sigma )\geq 0\ \ \mbox{\rm for \ all \ separable \ states}\ \ \sigma \ {\rm
on}\  H\otimes K.$$}

{\bf Proof.} Still, we only need to prove the ``only if" part.
Assume that $\rho$ is entangled. Then, by Lemma 2.1,  there exists a
real number $\alpha_0\in{\mathbb R}$ and a self-adjoint operator
$T\in{\mathcal T}(H\otimes K)$ such that the operator $W_0=\alpha_0
I+T$ satisfies ${\rm Tr}(W_0\rho)<0$ and ${\rm Tr}(W_0\sigma )\geq
0$ for all separable states $\sigma$ on $H\otimes K$. For such $T$,
by spectral theorem, there exists an orthogonal set
$\{|\omega_i\rangle\langle\omega_i|\}_{i=1}^k$ of rank one
projections on $H\otimes K$
 such that
$T=\sum_{i=1}^k\alpha_i|\omega_i\rangle\langle\omega_i|$, where
$\sum_{i=1}^k|\alpha_i|<\infty$ and the index is arranged so that
$|\alpha_i|\geq|\alpha_{i+1}|$ for each  $i$. If $k<\infty$, let
$\alpha=\alpha_0$ and $R=T$. If $k=\infty$, that is,
$T=\sum_{i=1}^\infty\alpha_i|\omega_i\rangle\langle\omega_i|$, since
${\rm Tr}(W_0\rho)<0$, there exists a positive number $\varepsilon$
such that ${\rm Tr}(W_0\rho)\leq-\varepsilon<0$. Let
$W_\varepsilon=W_0+\frac{1}{2}\varepsilon I$. Then we have ${\rm
Tr}(W_\varepsilon\sigma )\geq\frac{1}{2}\varepsilon> 0$ holds for
all $\sigma\in{\mathcal S}_{\rm sep}$ and  $${\rm
Tr}(W_\varepsilon\rho)=\alpha_0+\frac{1}{2}\varepsilon+\sum_{i=1}^\infty\alpha_i{\rm
Tr}(|\omega_i\rangle\langle\omega_i|\rho)\leq-\frac{1}{2}\varepsilon<0.$$

Let
$A_n=(\alpha_0+\frac{1}{2}\varepsilon)I+\sum_{i=1}^n\alpha_i|\omega_i\rangle\langle\omega_i|$.
Since $\|A_n-W_\varepsilon\|=|\alpha_{n+1}|\rightarrow 0$ as
$n\rightarrow\infty$, there exists a natural number $N_1$ such that
$\|A_n-W_\varepsilon\|<\frac{1}{4}\varepsilon$ whenever $n\geq N_1$.
Now, for any separable state $\sigma$,
$$\begin{array}{rl}{\rm Tr}(A_n\sigma)\geq&
{\rm Tr}(W_\varepsilon\sigma)-|{\rm Tr}(A_n\sigma)-{\rm
Tr}(W_\varepsilon\sigma)|\\
 \geq&{\rm
Tr}(W_\varepsilon\sigma)-\|A_n-W_\varepsilon\|\geq
\frac{1}{2}\varepsilon-\frac{1}{4}\varepsilon=\frac{1}{4}\varepsilon>0.\end{array}$$
Therefore $A_n$ is an EW whenever $n\geq N_1$.

Since $\lim_{n\rightarrow\infty}\sum_{i=1}^n\alpha_i{\rm
Tr}(|\omega_i\rangle\langle\omega_i|\rho)=\sum_{i=1}^\infty\alpha_i{\rm
Tr}(|\omega_i\rangle\langle\omega_i|\rho)$,  for the above
$\varepsilon$, there exists a natural number $N_2$ such that
$$\sum_{i=1}^n\alpha_i{\rm Tr}(|\omega_i\rangle\langle\omega_i|\rho)<\sum_{i=1}^\infty\alpha_i{\rm
Tr}(|\omega_i\rangle\langle\omega_i|\rho)+\frac{1}{4}\varepsilon$$
whenever $n>N_2$. Let $N=\max\{N_1, N_2\}$. If $n\geq N$, we have
$$\begin{array}{rl}{\rm
Tr}(A_n\rho)=&\alpha_0+\frac{1}{2}\varepsilon+\sum_{i=1}^{n}\alpha_i{\rm
Tr}(|\omega_i\rangle\langle\omega_i|\rho)\\
<&\alpha_0+\frac{1}{2}\varepsilon+\sum_{i=1}^\infty\alpha_i{\rm
Tr}(|\omega_i\rangle\langle\omega_i|\rho)+\frac{1}{4}\varepsilon\\
<&-\frac{1}{2}\varepsilon+\frac{1}{4}\varepsilon=-\frac{1}{4}\varepsilon<0.\end{array}$$
Hence $A_n$ is an entanglement witness for $\rho$. Now let
$\alpha=\alpha_0+\frac{1}{2}\varepsilon$ and
$R=\sum_{i=1}^n\alpha_i|\omega_i\rangle\langle\omega_i|$.

So far, we have proved that, for entangled state $\rho$, there
exists a real number $\alpha\in{\mathbb R}$ and a finite rank
self-adjoint operator $R\in{\mathcal T}(H\otimes K)$ such that the
operator $W=\alpha I+R$ satisfies ${\rm Tr}(W\rho)<0$ and ${\rm
Tr}(W\sigma )\geq 0$ for all separable states $\sigma$ on $H\otimes
K$. To complete the proof of the theorem, we still need to show that
$\alpha\geq 0$. We claim that
$$\inf _{\sigma \in{\mathcal S}_{\rm sep}}{\rm Tr}(R\sigma)=0.$$
As $R\in{\mathcal B}(H\otimes K)$ is a finite rank self-adjoint
operator, there exist vectors $|\psi_k\rangle\in H\otimes K$ and
real numbers $b_k$, $k=1,2,\ldots ,n$, such that
$R=\sum_{k=1}^nb_k|\psi_k\rangle\langle\psi_k|$. Take any
orthonormal bases $\{|i\rangle\}$ and $\{|j\rangle\}$ in $H$ and
$K$, respectively. Then $\{|ij\rangle\}$ is a product basis of
$H\otimes K$. For any $|ij\rangle$, let
$\sigma_{ij}=|ij\rangle\langle ij|$. Then we have
$${\rm Tr}(R\sigma_{ij})=\sum_{k=1}^nb_k|\langle \psi_k
|ij\rangle|^2 .$$ Thus $\lim_{i,j\rightarrow\infty}{\rm
Tr}(R\sigma_{ij})=0$, this ensures that $\inf _{\sigma \in{\mathcal
S}_{\rm sep}}{\rm Tr}(R\sigma)=0$.  Hence $\alpha\geq 0$ as $\alpha
+{\rm Tr}(R\sigma)\geq 0$ for all $\sigma\in{\mathcal S}_{\rm sep}$,
 completing the proof.
\hfill$\square$

\section{Constructing entanglement witnesses}

By Theorem 2.2, for any entangled state one may find an entanglement
witness of the form $\alpha I-T$ for it, where $\alpha\geq 0$ and
$T$ is a finite rank self-adjoint operator. If $\alpha\not=0$, this
witness may be replaced by $I-F$ with $F=\alpha^{-1}T$. In this
section, we will provide a method of constructing entanglement
witnesses of the such form, which enables us to obtain some examples
of entangled states that cannot be detected by the PPT criterion and
the realignment criterion.

{\bf 3.1. A Special Approach.} Let $H$ and $K$ be complex separable
Hilbert spaces and let $T$ be a self-adjoint operator on $H\otimes
K$. It is a natural idea that, if we find
$d_T=\sup_{\sigma\in{\mathcal S}_{\rm sep}}|{\rm Tr}(T\sigma)|$,
where ${\mathcal S}_{\rm sep}$ is the set of all separable states in
the composite system on $H\otimes K$, then $d_TI-T$ is an
entanglement witness if $d_T I-T$ is not positive. Once $d_T<\|T\|$,
the operator norm of $T$ (namely,
$\|T\|=\sup_{\||\omega\rangle\|=1}\|T|\omega\rangle\|$), then, for
any $d\in [d_T,\|T\|)$, $W=dI-T$ is an entanglement witness for some
entangled states. Thus, it is important to make an estimate of
$d_T=\sup_{\sigma\in{\mathcal S}_{\rm sep}}|{\rm Tr}(T\sigma)|$. In
this subsection we give a method to find such number $d_T$ and
construct entanglement witness of the form $d_TI-T$. Some examples
are also given to illustrate how to use this method to construct
entanglement witnesses.

In the following, we always assume that $H$ and $K$ are complex
Hilbert spaces that may be infinite-dimensional. $\{|i\rangle:
i\in\Lambda\}$  and $\{|j\rangle: j\in\Gamma\}$ be any fixed
orthonormal bases  of $H$ and  $K$, respectively.  Then every vector
$|\omega\rangle\in H\otimes K$ can be written in
$|\omega\rangle=\sum_{ij}d_{ij}|ij\rangle$ with
$\sum_{ij}|d_{ij}|^2=\||\omega\rangle\|^2<\infty$. The coefficient
matrix $D_\omega=(d_{ij})$ may be regarded as a Hilbert-Schmidt
operator from $K$ into $H$ defined by $D_{\omega}
|j\rangle=\sum_{i}d_{ij}|i\rangle$. Thus $D_{\omega}$ is a bounded
operator and we will denote its operator norm by $\|D_{\omega}\|$.

By Theorem 2.2, to construct an entanglement witness $W$, we may
assume that $W$ has the form $W=\alpha I-T$ with $T$ a self-adjoint
finite rank  operator. Thus
$T=\sum_{k}\alpha_k|\omega_k\rangle\langle\omega_k|$ for some
orthonormal set $\{|\omega_k\rangle\}_k\subset H\otimes K$ and some
real numbers $\alpha_k$ with $\|T\|_1={\rm
Tr}(|T|)=\sum_k|\alpha_k|$, where $|T|=(T^\dagger T)^{\frac{1}{2}}$.
Write $|\omega_k\rangle=\sum_{ij}d_{ij}^{(k)}|ij\rangle$. Regard
$D_k=(d_{ij}^{(k)})$ as operators from $K$ into $H$ and denote
$c_T=\sum_k |\alpha_k|\|D_k\|^2$.

{\bf Proposition 3.1.} {\it Let $\rho=p\rho_1+(1-p)\rho_2$ with
$0<p\leq 1$ be a state on $H\otimes K$. Then $W=c_{\rho_1}I-\rho_1$
is an entanglement witness if $W$ is not positive. Moreover, if
$$c_{\rho_1}<p\|\rho_1\|_2^2+(1-p){\rm Tr}(\rho_1\rho_2),$$
then  $\rho$ is entangled with $W$ its entanglement witness.}

{\bf Proof.} Write
$\rho_1=\sum_{k}\alpha_k|\omega_k\rangle\langle\omega_k|$. For any
unit product vector $|\mu\rangle|\nu\rangle\in H\otimes K$, we have
$$\begin{array}{rl}
|{\rm Tr}(\rho_1(|\mu\rangle |\nu\rangle\langle\mu |\langle\nu|))|
=&|\langle\mu|\langle\nu|\rho_1|\mu\rangle
|\nu\rangle|\\
\leq& \sum_k|\alpha_k||\langle\omega_k|\mu\rangle|\nu\rangle|^2 \\
=&\sum_k|\alpha_k||\langle\mu|D_k|\bar{\nu}\rangle|^2\\
\leq&\sum_k |\alpha_k|\|{D_k}\|^2=c_{\rho_1}.\end{array}$$  Let
$W=c_{\rho_1}I-\rho_1$. It follows  that ${\rm Tr}(W\sigma)\geq 0$
for all separable pure states $\sigma$ on $H\otimes K$. Thus $W$ is
an entanglement witness whenever $W$ is not positive.

If $c_{\rho_1}<p\|\rho_1\|_2^2+(1-p){\rm Tr}(\rho_1\rho_2),$ then
$${\rm Tr}(W\rho)=c_{\rho_1}-p\|\rho_1\|_2^2-(1-p){\rm
Tr}(\rho_1\rho_2)<0.$$ Hence, by Lemma 2.1, we have that $\rho$ is
entangled with entanglement witness $W$. \hfill$\Box$

Particularly, we have

{\bf Corollary 3.2.} {\it Let
$\rho=\sum_{k}p_k|\omega_k\rangle\langle\omega_k|$ be a state acting
on $H\otimes K$ with
$\{|\omega_k\rangle=\sum_{ij}d_{ij}^{(k)}|ij\rangle\}_k$ an
orthonormal set in $H\otimes K$. Regard $D_k=(d_{ij}^{(k)})$ as
operators from $K$ into $H$. If there is some $k_0$ such that
$\|D_{k_0}\|^2<p_{k_0}$, then $\rho$ is an entangled state and
$\|D_{k_0}\|^2I-|\omega_{k_0}\rangle\langle\omega_{k_0}|$ is an
entanglement witness for $\rho$. }

{\bf Proof.} Let
$\rho_{k_0}=\frac{1}{1-p_{k_0}}\sum_{l\not=k_0}p_l|\omega_l\rangle\langle\omega_l|$.
Then $\rho=p_{k_0}|\omega_{k_0}\rangle\langle\omega_{k_0}|
+(1-p_{k_0})\rho_{k_0}$. Note that, ${\rm
Tr}(|\omega_{k_0}\rangle\langle\omega_{k_0}|\rho_{k_0})=0$. So, by
Proposition 3.1,
$W=\|D_{k_0}\|^2I-|\omega_{k_0}\rangle\langle\omega_{k_0}|$ is an
entanglement witness if $W$ is not positive, and if
$\|D_{k_0}\|^2<p_{k_0}$, then ${\rm
Tr}(W\rho)=\|D_{k_0}\|^2-p_{k_0}<0$, that is, $\rho$ is
entangled.\hfill$\Box$

Let us consider the special case that
$\rho=|\psi\rangle\langle\psi|$ is a pure state. By Corollary 3.2,
if $\|D_{\psi}\|<1$, then $W=\|D_\psi\|^2I-|\psi\rangle\langle\psi|$
is an entanglement witness detecting $\rho$. This case was discussed
 in \cite{BE} for finite-dimensional systems and got the same
 conclusion.

 In the following, we give some examples of  entangled
states that can be detected by  the entanglement witness $W$
constructed in Proposition  3.1 or Corollary 3.2.

 {\bf Example 3.3.} Let $H$ and $K$ be
infinite-dimensional complex separable Hilbert spaces with
orthonormal basis $\{|i\rangle\}_{i=0}^\infty$ and
$\{|j\rangle\}_{j=0}^\infty$, respectively. Let
$$\rho=p_1|\omega_1\rangle\langle\omega_1|+p_2|\omega_2\rangle\langle\omega_2|
+p_3|\omega_3\rangle\langle\omega_3|$$ with $p_k\geq 0$ and
$p_1+p_2+p_3=1$, where
$$|\omega_1\rangle=\sqrt{\frac{6}{\pi^2}}\sum_{i=0}^\infty\frac{1}{i+1}|ii\rangle, \ \ \  \
|\omega_2\rangle=\sqrt{\frac{6}{\pi^2}}\sum_{i=0}^\infty\frac{1}{i+1}|(i+1)i\rangle$$
and
$$|\omega_3\rangle=\sqrt{\frac{6}{\pi^2}}\sum_{i=0}^\infty\frac{1}{i+1}|(i+2)i\rangle.$$
Then, $\rho$ is a state acting on $H\otimes K$ and
$\{\omega_1,\omega_2,\omega_3\}$ is an orthonormal set.  It is clear
that, regarded as operators from $K$ into $H$, $D_1$, $D_2$ and
$D_3$ as that in Corollary 3.2 are determined by
$D_1|i\rangle=\frac{\sqrt{6}}{\pi (i+1)}|i\rangle$,
$D_2|i\rangle=\frac{\sqrt{6}}{\pi (i+1)}|i+1\rangle$ and
$D_3|i\rangle=\frac{\sqrt{6}}{\pi (i+1)}|i+2\rangle$, respectively.

 Since $\|D_k\|^2=\frac{6}{\pi^2}$, $k=1,2,3$,
so, by use of Corollary 3.2, one sees that, if $p_k>\frac{6}{\pi^2}$
for some $k=1,2,3$, then $\rho$ is entangled with entanglement
witness $W=\frac{6}{\pi^2}I-|\omega_k\rangle\langle\omega_k|$.

{\bf Example 3.4.} Let $H$ and $K$ be $n$-dimensional complex
Hilbert spaces with   orthonormal bases $\{|i\rangle\}_{i=0}^{n-1}$
and $\{|j\rangle\}_{j=0}^{n-1}$, respectively. Let
$$|\omega_{1}\rangle=\frac{1}{\sqrt{n}}(|00\rangle+|11\rangle+\ldots
+|(n-1)(n-1)\rangle),$$
$$|\omega_{2}\rangle=\frac{1}{\sqrt{n}}(|01\rangle+|12\rangle+\ldots
+|(n-1)0\rangle),$$
$$\vdots $$
$$|\omega_{n}\rangle=\frac{1}{\sqrt{n}}(|0(n-1)\rangle+|10\rangle+\ldots
+|(n-1)(n-2)\rangle).$$ Let $\rho=\sum_{i=1}^nq_i\rho_i$ with
$\rho_i=|\omega_i\rangle\langle\omega_i|$, where $q_i\geq 0$ for
$i=1,2,\ldots ,n$ and $\sum_{i=1}^nq_i=1$. Then $\rho$ is a state on
$H\otimes K$. If $q_i$ ($i=1,2,\ldots ,n$) are not the same,
 then there must be  some
$i_0\in\{1,2,\ldots, n\}$ such that $q_{i_0}>\frac{1}{n}$. Without
loss of generality, assume that $q_{1}>\frac{1}{n}$. Let
$W=\frac{1}{n}I-\rho_{1}$. By Corollary 3.2, ${\rm
Tr}(W\sigma)\geq0$ for all separable states $\sigma$ and ${\rm
Tr}(W\rho)<0$, which imply that $\rho$ is entangled with
entanglement witness $W$.

If $q_1=q_2=\ldots =q_n=\frac{1}{n}$, then $\rho$ is PPT. Moreover,
by \cite[Theorem 1]{WCZ},
 $\rho$ is separable.

{\bf 3.2. A General Approach.} In the following, we will consider
more general cases and provide a method of constructing entanglement
witnesses. This method is motivated by an idea used in \cite{JB}.

Let $\rho$ be a mixed state. Write $\rho$ in the form
$\rho=\sum_ip_i\rho_i$, where $\rho_i$'s are pure states with
$\rho_i\rho_j=0$ whenever $i\not=j$. By Theorem 2.2, it is possible
to construct an entanglement witness on $H\otimes K$ of the form
$W=\alpha I+\sum_{i=1}^na_i \rho_i$ that may detect the entanglement
of $\rho$.

To do this, let $L:{\mathcal T}_{\rm sa}(H\otimes K)\rightarrow
{\mathbb R}^n$ be a map defined by $$L(T)=({\rm Tr}(\rho_1T),{\rm
Tr}(\rho_2T),\ldots,{\rm Tr}(\rho_nT))^t$$ for every $T\in{\mathcal
T}_{\rm sa}(H\otimes K)$, the real linear space of all self-adjoint
operators in ${\mathcal T}(H\otimes K)$. It is obvious that $L$ is
linear, bounded and $L({\mathcal S}_{\rm sep})\subseteq{\mathbb
R}^n$ is a closed convex set. If $L(\rho)\not\in L({\mathcal S}_{\rm
sep})$, then $\rho$ is entangled and we can find a linear functional
$f=(a_1,a_2,\ldots,a_n)\in({\mathbb R}^n)^*$ such that
$$f(L(\rho))>1 \ \ {\rm and}\ \  L({\mathcal S}_{\rm sep})\subseteq\{|x\rangle\in{\mathbb R}^n: f(|x\rangle)\leq 1\}$$
or
$$f(L(\rho))<1 \ \ {\rm and}\ \  L({\mathcal S}_{\rm sep})\subseteq\{|x\rangle\in{\mathbb R}^n: f(|x\rangle)\geq 1\}.$$
 Let $W=I-\sum_{i=1}^na_i \rho_i$ or $W=-I+\sum_{i=1}^na_i \rho_i$. Then $W$ satisfies ${\rm
Tr}(W\sigma )\geq 0$ for all separable states $\sigma$ on $H\otimes
K$ and ${\rm Tr}(W\rho)<0$, that is, $W$ is an entanglement witness
that can detect $\rho$. In fact, by Theorem 2.2, the second case
never occurs. So, we find a witness of the form $W=I-\sum_{i=1}^na_i
\rho_i$ that detects some entangled state of the form
$\rho=\sum_ip_i\rho_i$. Note that, the special approach in
Subsection 3.1 is the case of taking $n=1$.

One of the difficulties of the above general approach is that, in
general, the map $L$ has too many even infinite many variables. In
the case that there exist finite rank projections $P\in{\mathcal
B}(H)$ and $Q\in{\mathcal B}(K)$ such that $\rho_k=(P\otimes
Q)\rho_k(P\otimes Q)$, the situation may be simplified. In this
case, denote by ${\mathcal B}_{\rm sa}(PH\otimes QK)$ the space of
all self-adjoint operators acting on the finite dimensional space
$PH\otimes QK$. Let $L^\prime=L|_{{\mathcal B}_{\rm sa}(PH\otimes
QK)}$ and
 $\rho^\prime =\frac{1}{p_1+\ldots
+p_n}\sum_{k=1}^np_k\rho_k|_{PH\otimes QK}\in {\mathcal S}(PH\otimes
QK)$. If $L^\prime (\rho^\prime)\not\in L^\prime ({\mathcal S}_{\rm
sep}(PH\otimes QK))$, then $\rho^\prime$ is entangled with a witness
of the form $W^\prime =I_{PH\otimes
QK}-\sum_{k=1}^na_k\rho_k|_{PH\otimes QK}$. We claim that
$W=I-\sum_{k=1}^na_k\rho_k$ is an entanglement witness for original
system $H\otimes K$. Take any orthonormal bases
$\{|i\rangle\}_{i=0}^l$ and $\{|j\rangle\}_{j=0}^m$ of $PH$ and
$QK$,  and extend them to orthonormal bases $\{|i\rangle\}$ and
$\{|j\rangle\}$ of $H$ and $K$, respectively. For any unit vectors
$|\psi\rangle\in H$ and $|\phi\rangle\in K$, we can write
$|\psi\rangle =\sum_{i}\xi_i|i\rangle$ and
$|\phi\rangle=\sum_j\eta_j|j\rangle$. Let $\sigma
=|\psi\rangle|\phi\rangle\langle\psi|\langle\phi|$ be the separable
pure state. Then
$$\begin{array}{rl}{\rm Tr}(W\sigma)=&1-\sum_{k=1}^n a_k{\rm
Tr}(\rho_k\sigma) \\ =&1-\sum_{k=1}^n a_k{\rm Tr}((PH\otimes
QK)\rho_k(PH\otimes QK)\sigma(PH\otimes QK))\\
=&{\rm Tr}(W^\prime (PH\otimes QK)\sigma(PH\otimes QK))\geq
0\end{array}
$$
since $(PH\otimes QK)\sigma(PH\otimes QK)$ is separable. Now it is
clear that $W$ is an entanglement witness as it is not positive.
Note that
$${\rm Tr}(W\rho)=(p_1+\ldots +p_n){\rm Tr}(W^\prime \rho^\prime)+\sum_{k>n}p_k. $$
So, $\rho$ is entangled if $\frac{1-p_1-\ldots -p_n}{p_1+\ldots
+p_n}<-{\rm Tr}(W^\prime \rho^\prime)$.

To illustrate how to use the general approach to construct
entanglement witnesses, we give an example here.

{\bf Example 3.5.} Let $H$ and $K$ be  complex Hilbert spaces with
orthonormal bases $\{|i\rangle\}_{i=0}^\infty$ and
$\{|j\rangle\}_{j=0}^\infty$, respectively. Let
$$|\omega_{1}\rangle=\frac{1}{\sqrt{3}}(|00\rangle+|11\rangle+|22\rangle)$$
and
$$|\omega_{2}\rangle=\frac{1}{\sqrt{3}}(|01\rangle+|12\rangle+|20\rangle).$$
Define $\rho_1=|\omega_1\rangle\langle\omega_1|$,
$\rho_2=|\omega_2\rangle\langle\omega_2|$ and
$$\rho_3=\frac{1}{3}(|02\rangle\langle02|
+|10\rangle\langle10|+|21\rangle\langle21|).$$ Let $\rho_4$ be any
state satisfying ${\rm Tr}(\rho_k\rho_4)=0$ for $k=1,2,3$. Then, for
any $t\in(0,1)$,
$$\tilde{\rho}_t=(1-t)\sum_{i=1}^3q_i\rho_i+t\rho_4$$ is a state on
$H\otimes K$, where $q_i\geq 0$ for $i=1,2,3$ and $q_1+q_2+q_3=1$.
Let us discuss the entanglement of $\tilde{\rho}_t$.

Let $H_1$ and $K_1$ be the 3-dimensional spaces spanned by
$|0\rangle,|1\rangle, |2\rangle$ respectively. Thus $\rho_i$,
$i=1,2,3$, defined above can be regarded as states on $H_1\otimes
K_1$. Let  $\rho=\sum_{i=1}^3q_i\rho_i\in {\mathcal S}(H_1\otimes
K_1)$. By the general approach stated above, if we may construct
some entanglement witnesses of the form $W^\prime =I_{3\times
3}-(a_1\rho_1+a_2\rho_2+a_3\rho_3)|_{H_1\otimes K_1}$ for the
$3\times 3$ system, then we get
 some entanglement    witnesses of the same form
$W=I-(a_1\rho_1+a_2\rho_2+a_3\rho_3)$, which may detect entangled
states $\tilde{\rho}_t$ for suitable $q_1,q_2, q_3$ and $t$. In
fact,
 we will find two entanglement witnesses
$W_1=I-1.5\rho_1-0.3\rho_2-3\rho_3$ and
$W_2=I-0.3\rho_1-1.5\rho_2-3\rho_3$, which detect $\rho$ for some
suitable $q_1,q_2, q_3$, and then detect $\tilde{\rho}_t$ for
sufficient small $t$; particularly, if $q_2<\frac{5}{7}q_1$ or
$q_1<\frac{5}{7}q_2$, and if $q_1q_2q_3\geq q_1^3+q_2^3$, then
$\tilde{\rho}_t$ is an entangled PPT state whenever  $\rho_4$ is.

The realignment criterion can be generalized to infinite dimensional
systems. For a pure state $\rho=|\psi\rangle\langle\psi|$ with
$|\psi\rangle=\sum\limits_{i,j}d_{ij}|ij\rangle$, writing
$D=(d_{ij})$ and $\bar{D}=(\bar{d}_{ij})$, the realignment operator
of $\rho$ is defined by $\rho^R=D\otimes \bar{D}$. Furthermore, if
$\rho=\sum\limits_ip_i\rho_i$ is a mixed state,
 where $\sum\limits_ip_i=1$, $p_i>0$,
$\rho_i$'s are pure states, then  the realigned operator of $\rho$
is defined by $\rho^R=\sum\limits p_i\rho_i^R$ \cite{GQH}. Thus the
realignment criterion states that, for any state
$\rho\in\mathcal{S}(H\otimes K)$, if $\rho$ is separable, then
$\|\rho^{R}\|_{1}\leq 1$ \cite{GY}. Thus, applying the entanglement
witnesses $W_1$ and $W_2$, we see that, if
$0<2q_2=q_1\leq\frac{2}{303}$ or $0<2q_1=q_2\leq\frac{2}{303}$, and
if $\rho_4$ is separable and $t$ is small enough, then
 $\tilde{\rho}_t$ is  PPT entangled that cannot be detected by the
realignment criterion.

To show the desired conclusion, by the general approach and the
discussion above, in the following, we need only regard $\rho$ and
$\rho_i$, $i=1,2,3$, as states of $3\times 3$ system $H_1\otimes
K_1$ and show that $W_1=I-1.5\rho_1-0.3\rho_2-3\rho_3$ and
$W_2=I-0.3\rho_1-1.5\rho_2-3\rho_3$ are entanglement witnesses, and
then
 apply them to
check that

(1) if $q_2<\frac{5}{7}q_1$ or $q_1<\frac{5}{7}q_2$, then $\rho$ is
entangled;

(2) if $q_2<\frac{5}{7}q_1$ or $q_1<\frac{5}{7}q_2$, and if
$q_1q_2q_3\geq q_1^3+q_2^3$, then $\rho$ is PPT entangled;

(3) if $0<2q_2=q_1\leq\frac{2}{303}$ or
$0<2q_1=q_2\leq\frac{2}{303}$, then $\rho$ is  PPT entangled that
cannot be detected by  the realignment criterion.

It is easily checked that $\rho$ is PPT if and only if
$q_1q_2q_3\geq q_1^3+q_2^3$. And by Corollary 3.2, if
$q_1>\frac{1}{3}$ or $q_2>\frac{1}{3}$, then $\rho$ is NPPT
entangled.

Since every separable state can be written as a  convex combination
of pure product states, to find an entanglement witness of the form
$W=I-a_1\rho_1-a_2\rho_2-a_3\rho_3$, we only need to consider the
values ${\rm
Tr}(\rho_i(|\alpha\rangle|\beta\rangle\langle\alpha|\langle\beta|))$
for all pure product states $|\alpha\rangle|\beta\rangle$ on
$H_1\otimes K_1$. Let
$|\alpha\rangle=(\alpha_1,\alpha_2,\alpha_3)^{\rm t}\in H_1$ and
$|\beta\rangle=(\beta_1,\beta_2,\beta_3)^{\rm t}\in K_1$ be any unit
vectors, and let
$$c_1={\rm
Tr}(\rho_{1}(|\alpha\rangle|\beta\rangle\langle\alpha|\langle\beta|))=\frac{1}{3}(\alpha_1\beta_1+\alpha_2\beta_2+\alpha_3\beta_3)^2,$$
$$c_2={\rm
Tr}(\rho_{2}(|\alpha\rangle|\beta\rangle\langle\alpha|\langle\beta|))=\frac{1}{3}(\alpha_1\beta_2+\alpha_2\beta_3+\alpha_3\beta_1)^2,$$
$$c_3={\rm
Tr}(\rho_3(|\alpha\rangle|\beta\rangle\langle\alpha|\langle\beta|))=\frac{1}{3}(\alpha_1^2\beta_3^2+\alpha_2^2\beta_1^2+\alpha_3^2\beta_2^2).$$
As the maximum value of $c_i$ ($i=1,2,3$) is $\frac{1}{3}$, the
points $A=(\frac{1}{3},0,0)$, $B=(0,\frac{1}{3},0)$ and
$C=(0,0,\frac{1}{3})$ are vertices of the convex set $L({\mathcal
S}_{\rm sep})$, where $L:{\mathcal T}_{\rm sa}(H_1\otimes
K_1)\rightarrow {\mathbb R}^{3}$ is the linear map defined by
$L(T)=({\rm Tr}(T\rho_1),{\rm Tr}(T\rho_2),{\rm Tr}(T\rho_3))^t$.
The plane which passes through $A,B,C$ has the following equation
$$3(x_1+x_2+x_3)=1.$$
Note that, if
$|\alpha\rangle=|\beta\rangle=(\frac{1}{\sqrt{3}},\frac{1}{\sqrt{3}},\frac{1}{\sqrt{3}})$,
then $c_1=c_2=\frac{1}{3}$ and $c_3=\frac{1}{9}$, which implies that
$3(c_1+c_2+c_3)=\frac{7}{3}>1$. Hence the point
$D=(\frac{1}{3},\frac{1}{3},\frac{1}{9})$ lies in the half-space
$3(x_1+x_2+x_3)>1$.

Now consider the plane
$$3x_1-x_2+3x_3=1$$
passing through $A,C$ and $D$. Pick
$|\alpha\rangle=|\beta\rangle=(\frac{1}{\sqrt{2}},\frac{1}{\sqrt{2}},0)$,
we have $c_1=\frac{1}{3}$ and $c_2=c_3=\frac{1}{12}$, which imply
that $3c_1-c_2+3c_3=\frac{13}{12}>1$. Hence the point
$E=(\frac{1}{3},\frac{1}{12},\frac{1}{12})\in L({\mathcal S}_{\rm
sep})$ is in the half-space $3x_1-x_2+3x_3>1$.

Next, let us   consider the plane passing through $A,  C$ and $E$,
which has the equation
$$3(x_1-x_2+x_3)=1.$$
If we take
$$|\alpha\rangle=(-0.707104,0.672502,-0.218509)$$ and
$$|\beta\rangle=(0.707107,-0.706824,-0.0199903),$$ we get $c_1=0.314261,$
$c_2=0.0367071$ and $c_3=0.0833943$. In this case,
$3(c_1-c_2+c_3)=1.08284>1$. Denote
$F=(0.314261,0.0367071,0.0833943)$. Then the plane passing through
the points $C, D$ and $F$ has the equation
$$2.43701x_1-0.43701x_2+3x_3=1.$$
Take $|\alpha\rangle=(0.876317,-0.0152726,0.481493)$ and
$|\beta\rangle=(0.481493,-0.0152726,0.876317)$. Then $(c_1,
c_2,c_3)=(0.237509,0.0140176,0.196609)$ and
$2.43701c_1-0.43701c_2+3c_3=1.16251>1$. Let $G=(0.237509, 0.0140176,
0.196609)$.

Now we consider the plane passing through the points $C$, $E$ and
$G$. The plane has the equation
$$1.71x_1+0.29x_2+3x_3=1.\eqno(3.1)$$
Since the maximum value of $1.71c_1+0.29c_2+3c_3$ is around
$1.0174$, which is much close to 1, we can make the plane (3.1) a
small rotation around the point $C$ to get a new plane that is
tangent to $L({\mathcal S}_{\rm sep})$. For instance, let us
consider the following one
$$1.5x_1+0.3x_2+3x_3=1.\eqno(3.2)$$
As the maximum  value of $1.5c_1+0.3c_2+3c_3$ is 1, so the plane
(3.2) is tangent to $L({\mathcal S}_{\rm sep})$  and
$$L({\mathcal S}_{\rm sep})\subset\{X=(x_1,x_2,x_3)^t :
1.5x_1+0.3x_2+3x_3\leq 1\}. \eqno(3.3)$$

Define $$W_1=I-1.5\rho_1-0.3\rho_2-3\rho_3.\eqno(3.4)$$ By Eq.(3.3),
we have ${\rm Tr}(W\sigma)\geq0$ for all separable states $\sigma$
on $H_1\otimes K_1$. Clearly, $W_1$ is an entanglement witness as it
is not positive.

Symmetrically, by replacing point $A$ by $B$, one gets an
entanglement witness of the form
$$W_2=I-0.3\rho_1-1.5\rho_2-3\rho_3.\eqno(3.5)$$

Note that, by Eq.(3.4),
$${\rm Tr}(W_1\rho)=1-1.5q_1-0.3q_2-q_3=-0.5q_1+0.7q_2.\eqno(3.6)$$
So the entanglement of $\rho$ can be detected by the entanglement
witness $W_1$ if $0\leq q_2<\frac{5}{7}q_1$. Furthermore, if
$q_1,q_2,q_3$ satisfy $q_1q_2q_3\geq q_1^3+q_2^3$ and $0\leq
q_2<\frac{5}{7}q_1$, then $\rho$ is a PPT entangled state that can
be detected by $W_1$.

Now let's consider the realignment criterion.

For $\rho=q_1\rho_1+q_2\rho_2+q_3\rho_3$, it is obvious that
$$\rho=\left(\begin{array}{ccccccccc}q_1&0&0&0&q_1&0&0&0&q_1\\
0&q_3&0&0&0&0&0&0&0\\
0&0&q_2&q_2&0&0&0&q_2&0\\
0&0&q_2&q_2&0&0&0&q_2&0\\
q_1&0&0&0&q_1&0&0&0&q_1\\
0&0&0&0&0&q_3&0&0&0\\
0&0&0&0&0&0&q_3&0&0\\
0&0&q_2&q_2&0&0&0&q_2&0\\
q_1&0&0&0&q_1&0&0&0&q_1
\end{array}\right). $$
Then the realignment matrix of $\rho$ is
$$\rho^R=\left(\begin{array}{ccccccccc}q_1&0&0&0&q_3&0&0&0&q_2\\
0&q_1&0&0&0&0&q_2&0&0\\
0&0&q_1&0&0&0&0&q_2&0\\
0&0&q_2&q_1&0&0&0&0&0\\
q_2&0&0&0&q_1&0&0&0&q_3\\
0&q_2&0&0&0&q_1&0&0&0\\
0&0&0&0&0&q_2&q_1&0&0\\
0&0&0&q_2&0&0&0&q_1&0\\
q_3&0&0&0&q_2&0&0&0&q_1
\end{array}\right). $$
By computation, we have that the trace norm $\|\rho^R\|_1<1$ if
$q_1\leq\frac{2}{303}$ and $q_2=\frac{1}{2}q_1$. Hence, in this
case, the entangled state $\rho$ is PPT and cannot be detected by
the realignment criterion. however it can be detected by the
entanglement witness $W_1$. Applying the entanglement witness $W_2$
in Eq.(3.5), one gets the other half of the assertions (1)-(3).

{\bf Conclusion.} For infinite-dimensional composite systems, every
entangled state   has a simple entanglement witness of the form
$\alpha I-T$,  where   $\alpha $ is a nonnegative number and $T$ is
a self-adjoint  operator of finite rank. Such kind of witnesses have
nice structure and can be handled easily. Based on this fact, two
approaches of constructing entanglement
 witnesses of such form for entangled states are provided. Some  examples of entanglement witnesses of such form are
 found and used to detect entangled states
 that cannot be detected by the PPT criterion and the realignment
 criterion.

{\bf Acknowledgement.} The authors wish to give their thanks to the
referees for helpful comments to improve this paper.

\end{document}